\documentclass{ws-xxx-notrim}

\begin{document}


\title{Neutrino oscillations, and the origin of pulsar
  velocities and dark matter} 

\author{Alexander Kusenko\footnote{\uppercase{W}ork supported in part by
the \uppercase{DOE} grant \uppercase{DE-FG03-91ER40662} and the
\uppercase{NASA ATP} grant \uppercase{NAG5-13399}.}}

\address{Department of Physics and Astronomy, UCLA, Los Angeles, CA
90095-1547\\ RIKEN BNL Research Center, Brookhaven National Laboratory,
Upton, NY 11973 }

\maketitle

\abstracts{ Neutrino oscillations in a core-collapse supernova may be
responsible for the observed rapid motions of pulsars.  Three-dimensional
numerical calculations show that, in the absence of neutrino oscillations,
the recoil velocities of neutron stars should not exceed 200~km/s, while
there exists a substantial population of pulsars that move faster than
1000~km/s.  A small asymmetry in the neutrino emission may be the solution
of this long-standing puzzle.  Such an asymmetry could arise from neutrino
oscillations, but, given the present bounds on the neutrino masses, the
pulsar kicks require a sterile neutrino with a 1--20~keV mass and a small
mixing with active neutrinos.  The same particle can be the cosmological
dark matter.  Its existence can be confirmed by X-ray telescopes if they
detect X-ray photons from the decays of the relic sterile neutrinos.  One
can also verify the neutrino kick mechanism by observing gravity waves from
a nearby supernova.}

%
%
%
%

\section{Introduction}

Neutrino oscillations in a core-collapse supernova have been the subject of
intense studies.  It has been pointed out, in particular, that neutrino
oscillations may be the cause of otherwise unexplained large pulsar
velocities\cite{ks96,ks97}.  This explanation requires the existence 
a sterile neutrino\cite{ks97,fkmp} with mass and mixing that are
just right for the same particle to be the cosmological dark matter.  

Pulsar velocities range from 100 to 1600~km/s\cite{astro,astro_1}. Their
distribution is non-gaussian, with about 15\% of all pulsars having speeds
over (1000~km/s)\cite{astro_1}.
Pulsars are born in supernova explosions, so it would be natural to look
for an explanation in the dynamics of the
supernova\cite{explosion}. However, state-of-the-art 3-dimensional
numerical calculations\cite{fryer} show that even the most extreme
asymmetric explosions do not produce pulsar velocities greater than
200~km/s.  Earlier 2-dimensional calculations\cite{sn_2D} claimed a
somewhat higher maximal pulsar velocity, up to 500~km/s.  Of course, even
that was way too small to explain the large population of
pulsars whose speeds exceed 1000~km/s.  Recent three-dimensional
calculations by Fryer\cite{fryer} show an even stronger discrepancy. 

\section{Why a sterile neutrino can give the pulsar a kick}

Given the absence of a ``standard'' explanation, one is compelled to
consider alternatives, possibly involving new physics.  One of the reasons
why the standard explanation fails is because most of the energy is carried
away by neutrinos, which escape isotropically.  The remaining momentum must
be distributed with a substantial asymmetry to account for the large pulsar
kick.  In contrast, only a few per cent anisotropy in the distribution of
neutrinos, would give the pulsar a kick of required magnitude.

Neutrinos are always {\em \sf produced } with an asymmetry, but they 
{\em \sf escape} isotropically.  The asymmetry in production comes from the
asymmetry in the basic weak interactions in the presence of a strong
magnetic field.  Indeed, if the electrons and other fermions are polarized
by the magnetic field, the cross section of the urca processes, such as
$n+e^+ \rightleftharpoons p+ \bar \nu_e$ and $p+e^-\rightleftharpoons n+
\nu_e $, depends on the orientation of the neutrino momentum.  Depending on
the fraction of the electrons in the lowest Landau level, this asymmetry
can be as large as 30\%, much more than one needs to explain the pulsar
kicks\cite{drt}.  However, this asymmetry is completely washed out by
scattering of neutrinos on their way out of the star\cite{eq}.

If, however, the same interactions produced a particle which had {\em \sf even
weaker} interactions with nuclear matter than neutrinos, such a particle
could escape the star with an asymmetry equal its production asymmetry. 

It is intriguing that the same particle can the dark mater. 

The simplest realization of this scenario is a model that adds only one
singlet fermion to the Standard Model.  The SU(2)$\times$U(1) singlet, a
sterile neutrino, mixes with the usual neutrinos, for example, with the
electron neutrino.  

For a sufficiently small mixing angle between $\nu_e$ and $\nu_s$, only one
of the two mass eigenstates, $\nu_1$, is trapped.  The orthogonal state,
\begin{equation}
| \nu_2 \rangle = \cos \theta_m | \nu_s \rangle + \sin \theta_m | \nu_e
\rangle , 
\end{equation}
escapes from the star freely.  This state is produced in the same basic
urca reactions ($\nu_e+n\rightleftharpoons p+e^-$ and
$\bar\nu_e+p\rightleftharpoons n+e^+$) with the effective Lagrangian
coupling equal the weak coupling times $\sin \theta_m$.

We will consider two ranges of parameters, for which the $\nu_e \rightarrow
\nu_s$ oscillations occur on or off resonance. First, let us suppose that a 
resonant oscillation occurs somewhere in the core of the neutron star.
Then the asymmetry in the neutrino emission comes from shift in the
resonance point depending on the magnetic field\cite{ks97}.  Second, we
will consider the off-resonance case, in which the asymmetry comes directly
from the weak processes\cite{fkmp}.

\section{Resonant Mikheev-Smirnov-Wolfenstein oscillations}

Neutrino oscillations in a magnetized medium are described by an effective
potential\cite{magn}  
\begin{eqnarray}
V(\nu_{\rm s}) & = & 0  \label{Vnus} \\
V(\nu_{\rm e})& = & -V(\bar{\nu}_{\rm e}) =  V_0 \: (3 \, Y_e-1+4 \,
Y_{\nu_{\rm e}}) \label{Vnue} \\ 
V(\nu_{\mu,\tau}) & = & -V(\bar{\nu}_{\mu,\tau}) = V_0 \: ( Y_e-1+2 \, 
Y_{\nu_{\rm e}}) \ 
+\frac{e G_{_F}}{\sqrt{2}} \left ( \frac{3 N_e}{\pi^4} 
\right )^{1/3}
\frac{\vec{k} \cdot \vec{B}}{|\vec{k}|} \label{Vnumu}
\end{eqnarray}
where $Y_e$ ($Y_{\nu_{\rm e}}$) is the ratio of the number density of
electrons (neutrinos) to that of neutrons, $\vec{B}$ is the magnetic field,
$\vec{k}$ is the neutrino momentum, $V_0=10 \: \rm{eV} \: (\rho/10^{14} g
\, cm^{-3} )$.  The magnetic field dependent term in equation (\ref{Vnumu})
arises from polarization of electrons and {\em \sf not} from a neutrino
magnetic moment, which in the Standard Model is small and which we will
neglect. (A large neutrino magnetic moment can result in a pulsar kick
through a somewhat different mechanism\cite{voloshin}.)

The condition for resonant MSW oscillation $\nu_i \leftrightarrow
\nu_j$ is

\begin{equation}
\frac{m_i^2}{2 k} \: \cos \, 2\theta_{ij} + V(\nu_i) = 
\frac{m_j^2}{2 k} \: \cos \, 2\theta_{ij} + V(\nu_j)  
\label{res}
\end{equation}
where $\nu_{i,j}$ can be either a neutrino or an anti-neutrino. 

In the presence of the magnetic field, condition (\ref{res}) is
satisfied at different distances $r$ from the center, depending on the
value of the $(\vec{k} \cdot \vec{B})$ term in (\ref{res}). The average
momentum carried away by the neutrinos depends on the temperature of the
region from which they escape.  The deeper inside the star, the higher is
the temperature during the neutrino cooling phase.  Therefore, neutrinos
coming out in different directions carry momenta which depend on the
relative orientation of $\vec{k}$ and $\vec{B}$.  This causes the asymmetry
in the momentum distribution.

The surface of the resonance points is 

\begin{equation}
r(\phi) = r_0 + \delta \: cos \, \phi, 
\end{equation}
where $cos \, \phi= (\vec{k} \cdot \vec{B})/k$ and $\delta$ is determined
by the equation 
$(d N_n(r)/dr) \delta \approx 
e \left ( 3 N_e/\pi^4 \right )^{1/3} B$.
This yields\cite{ks97} 

\begin{equation}
\delta = 
\frac{e \mu_e}{ \pi^2} \: B \left / \frac{dN_n(r)}{dr} \right. ,
\label{delta}
\end{equation}
where $\mu_e \approx (3 \pi^2 N_e)^{1/3} $ is the chemical potential of the
degenerate (relativistic)  electron gas.

Assuming a black-body radiation luminosity $\propto T^4$, 
the asymmetry in the momentum distribution is\cite{ks97}
\begin{equation}
\frac{\Delta k}{k} = \frac{4 e}{3 \pi^2} \: \left ( \frac{\mu_e}{T}
\frac{dT}{dN_n} \right) B.
\label{dk1}
\end{equation}

To calculate the derivative in (\ref{dk1}), we use the relation between the
density and the temperature of a non-relativistic Fermi gas. Finally, 
\begin{equation}
\frac{\Delta k}{k} = \frac{4 e\sqrt{2}}{\pi^2} \: 
\frac{\mu_e \mu_n^{1/2}}{m_n^{3/2}T^2} \ B =
0.01 \left ( \frac{B}{3\times 10^{15} {\rm G} }\right )
\label{dk2}
\end{equation}
if the neutrino oscillations take place in the core of the neutron star, at
density of order $10^{14} \, {\rm g\,cm^{-3}}$.  The neutrino oscillations
take place at such a high density if one of the neutrinos has mass in the
keV range, while the other one is much lighter.  The magnetic field of the
order of $10^{15}-10^{16}$~G is quite possible inside a neutron star, where
it is expected to be higher than on the surface.  (In fact, some neutron
stars, dubbed magnetars, appear to have surface magnetic fields of this
magnitude.)

\begin{figure}[ht]
\centerline{\epsfxsize=4.1in\epsfbox{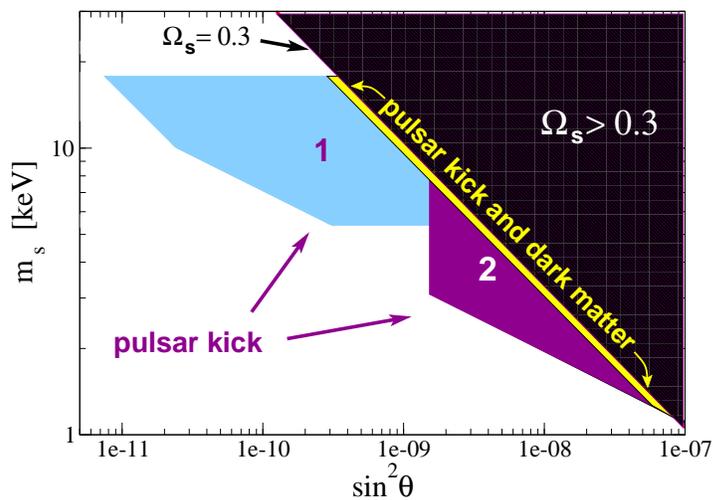}}   
\caption{The range of parameters for the sterile neutrino mass and mixing.
  Regions 1 and 2 correspond to parameters consistent with the pulsar kicks
  for (1) resonant and (2) off-resonant transitions, respectively.  Both
  regions overlap with a band in which the sterile neutrino is 
  dark matter.}
\label{figure:range}
\end{figure}

The region of parameters for which the asymmetric emission of sterile
neutrinos would result in a sufficient pulsar kick is shown in
Fig.~\ref{figure:range}.  Obviously, the mass has to be in the keV range
for the resonance to occur in the core of the neutron star.  Theoretical
models of neutrino masses can readily produce a sterile neutrino with
a required mass\cite{farzan,babu}.

Some comments are in order.  First, a similar kick mechanism, based
entirely on active neutrino oscillations (and no steriles) could also work
if the resonant oscillations took place between the electron and tau
neutrinospheres\cite{ks96,barkovich}.  This, however, would require the
mass difference between two neutrinos to be of the order of 100~eV, which
is ruled out.  Second, the neutrino kick mechanism was criticized
incorrectly by Janka and Raffelt\cite{jr}.  It was subsequently shown by
several authors\cite{ks98,barkovich} that Janka and Raffelt made several
mistakes, which is why their estimates were different from eq.~(\ref{dk2}).
In particular, Janka and Raffelt neglected the neutrino absorptions outside
the core, set the neutrino opacities to be equal to each other, regardless
of the neutrino flavor (which is incorrect\cite{ks98}), and neglected the
change in the neutrino flux due to the $1/r^2$ effect of the spherical
outflow\cite{barkovich}.

\section{Off-resonant oscillations}

For somewhat lighter masses, the resonant condition is not satisfied
anywhere in the core.  In this case, however, the off-resonant
production of sterile neutrinos in the core can occur through ordinary urca
processes.  A weak-eigenstate neutrino has a $\sin^2\theta $ admixture
of a heavy mass eigenstate $\nu_2$.  Hence, these heavy neutrinos can be
produced in weak processes with a cross section suppressed by $\sin^2\theta
$. 

Of course, the mixing angle in matter $\theta_m$ is not the same as it is
in vacuum, and initially $\sin^2\theta_m \ll \sin^2\theta$.  However, as
Abazajian, Fuller, and Patel\cite{Fuller} have pointed out, in the presence
of sterile neutrinos the mixing angle in matter quickly evolves toward its
vacuum value.  When $\sin^2\theta_m \approx \sin^2\theta$, the production 
of sterile neutrinos is no longer suppressed, and they can take a fraction
of energy out of a neutron star. 

Following Abazajian, Fuller, and Patel\cite{Fuller}, we have estimated the
time it takes for the matter potential to evolve to zero from its initial
value $V^{(0)}(\nu_e)\simeq (-0.2 ...+ 0.5) V_0$.  The time scale for this
change to occur through neutrino oscillations off-resonance is    
\begin{eqnarray}
\label{timeeqoffres}
 \tau_{_V}^\mathrm{off-res}  & \simeq &
\frac{4 \sqrt{2} \pi^2 m_n}{G_{\!\!_F}^3 \rho}
\frac{ (V^{(0)}(\nu_e))^3}{(\Delta m^2)^2 \sin^2 2 \theta } \frac{1}{\mu^3}
\\
& \sim & 
 \frac{6 \times 10^{-9} s}{\sin^2 2 \theta} 
\left (\frac{V^{(0)}(\nu_e)}{0.1 \mathrm{eV}} \right )^3 
\left (\frac{50 \mathrm{MeV}}{\mu} \right )^3 \left ( \frac{ 
10 \mathrm{keV}^2
}{\Delta m^2
} \right )^2. \nonumber   
\label{timeeqoffresnumerical}
\end{eqnarray}

As long as this time is much smaller than 10 seconds, the mixing angle in
matter approaches its value in vacuum in time for the sterile neutrinos to
take out some fraction of energy from a cooling neutron star.  

The urca processes produce ordinary neutrinos with some asymmetry depending
on the magnetic field\cite{drt}.  The same asymmetry is present in the
production cross sections of sterile neutrinos.  However, unlike the active
neutrinos, sterile neutrinos escape from the star without rescattering.
Therefore, the asymmetry in their emission is not washed out as it is in
the case of the active neutrinos\cite{eq}.  Instead, the asymmetry in
emission is equal the asymmetry in production.
 
\begin{figure}[ht]
\centerline{\epsfxsize=4.1in\epsfbox{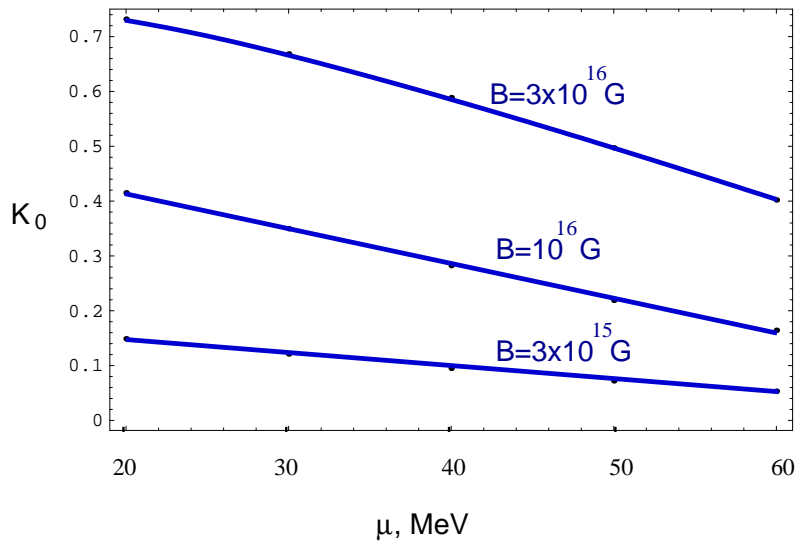}}   
\caption{The fraction of electrons in the lowest Landau level as a function
chemical potential.  
The value of the magnetic field in the core
of a neutron star is shown next to each curve.  }
\label{figure:asymmT20}
\end{figure}

The number
of neutrinos $dN$ emitted into a solid angle $d\Omega $ can be written as
\begin{equation}
\frac{dN}{d\Omega}= N_0(1+ \epsilon \cos \Theta_\nu ), 
\end{equation}
where $\Theta_\nu$ is the angle between the direction of the magnetic field
and the neutrino momentum, and $N_0$ is some normalization factor.  The
asymmetry parameter $\epsilon$ is equal
\begin{equation}
\epsilon = \frac{g_{_V}^2-g_{_A}^2}{g_{_V}^2+3g_{_A}^2} 
k_0   
\left ( \frac{  E_{\rm s}}{ E_{\rm tot}} \right ),  
\end{equation} 
where $g_{_V}$ and $g_{_A}$ are the axial and vector couplings, ${ E_{\rm
tot}}$ and $ { E_{\rm s}} $ are the total neutrino luminosity and the
luminosity in sterile neutrinos, respectively.  The number of electrons in
the lowest Landau level, $k_0$, depends on the magnetic field and the
chemical potential $\mu$ as shown in Fig.~\ref{figure:asymmT20}.

The momentum asymmetry in the neutrino emission is 
\begin{equation}
\epsilon \sim 0.02 
\left ( \frac{k_0}{0.3}\right ) 
\left ( \frac{ r_{_E}}{0.5} \right), 
\label{epsilon_final} 
\end{equation}  
where $r_{_E}$ is the fraction of energy carried by the sterile neutrinos.
To satisfy the constraint based on the observation of neutrinos 
from supernova SN1987A, we require that $r_{_E}<0.7$.  The asymmetry in
equation (\ref{epsilon_final}) 
can be of the order of the requisite few per cent for magnetic fields
$10^{15}-10^{16}$~G, as can be seen from Fig.~\ref{figure:asymmT20}.

Surface magnetic fields of pulsars are estimated to be of the order of
$10^{12}-10^{13}$G.  However, the magnetic field inside a neutron star may
be much higher\cite{magnetic,dt,magnetars}, probably up to $10^{16}$G.  The
existence of such a strong magnetic field is suggested by the dynamics of
formation of the neutron stars, as well as by the stability of the poloidal
magnetic field outside the pulsar\cite{dt}.  Moreover, the discovery of
soft gamma repeaters and their identification as
magnetars\cite{magnetars}, {\em i.e.}, neutron stars with {\em surface}
magnetic fields as large as $10^{15}$~G, gives one a strong reason to
believe that the interiors of many neutron stars may have magnetic fields
as large as $10^{15}-10^{16}$~G and that only in some cases this large
magnetic field breaks out to the surface.  There are also plausible
physical mechanisms capable of generating such a large magnetic field
inside a cooling neutron star\cite{magnetic,dt,zeldovich}.  

If the magnetic field inside a neutron star has a large non-dipole
component, the neutrino kick is off-centered.  Such a kick can probably
explain the unusually fast rotation of pulsars\cite{phinney}.

\section{Sterile neutrinos as dark matter; observational consequences}

Very few hints exist as to the nature of cosmological dark matter.  We know
that none of the Standard Model particles can be the dark matter, and we
also know that the dark matter particles should either be weakly
interacting or very heavy (or both).  Theoretical models have provided
plenty of candidates.  For example, the supersymmetric extensions of the
Standard Model predict the existence of a number of additional particles,
which include two dark matter candidates: the lightest supersymmetric
particle (LSP) and the SUSY Q-balls.  These are plausible candidates, which
were discussed in a number of talks at this conference.  In the absence of
observational hints, one naturally relies on theoretical models in planning
experimental searches for dark matter.

The parameter space allowed for the pulsar kicks\cite{fkmp} overlaps nicely
with that of dark-matter sterile neutrinos\cite{Fuller,dw}.  Sterile
neutrinos in this range may soon be discovered\cite{aft}.  Relic sterile
neutrinos with mass in the 1-20~keV range can decay into a lighter neutrino
and a photon.  The X-ray photons should be detectable by the X-ray
telescopes.  Chandra and XMM-Newton can exclude part of the parameter
space\cite{aft}.  The future Constellation-X can probably explore the
entire allowed range of parameters.

If inflation ended with a low-temperature reheating\cite{lowT}, the
allowed parameter space for the pulsar kicks extends to much lower masses
and larger mixing angles\cite{lowT}. 

In the event of a nearby supernova, the neutrino kick can produce gravity
waves that could be detected by LIGO and LISA\cite{loveridge,cuesta}. 

Active-to-sterile neutrino oscillations can give a neutron star a kick.
However, if a black hole is born in a supernova, it would not receive a
kick, unless it starts out as a neutron star and becomes a black hole
later, because of accretion. (The latter may be what happened in SN1987A,
which produced a burst of neutrinos, but no radio pulsar.) If the central
engines of the gamma-ray bursts are compact stars, the kick mechanism
acting selectively on neutron stars and not black holes could probably
explain the short bursts as interrupted long bursts\cite{k_semi}.

Since one does not expect a significant correlation between the 
magnetic field inside a hot neutron star (while this field is, presumably,
growing via the dynamo effect) and the eventual exterior field of a radio
pulsar, the neutrino kick mechanism does not predict any $B-v$ correlation. 

To summarize, asymmetric neutrino emission from a cooling neutron star can
explain the observed pulsar velocities.  The necessary condition for this
mechanism to work is the existence of a sterile neutrino with mass in the
1--20~keV range and a small mixing with ordinary neutrinos.  It is
intriguing that the same particle is a viable dark matter candidate.  The
nature of cosmological dark matter is still unknown.  We know that at least
one particle beyond the Standard Model must exist to account for dark
matter.  This particle may come as part of a ``package'', for example, if
supersymmetry is right.  However, it may be that the dark matter particle
is simply an SU(2)$\times$U(1) singlet fermion, which has a small mixing
with neutrinos.  Furture observations of X-ray telescopes have the
potential to discover the relic sterile neutrinos by detecting keV 
photons from the sterile neutrino decay in clusters of galaxies.  If
gravitational waves are detected from a nearby supernova, the signal may
show the signs of a neutron star being accelerated by an asymmetric
neutrino emission.


\begin{thebibliography}{0}

\bibitem{ks96}
  A.~Kusenko and G.~Segr\`e, Phys.~Rev.~Lett. {\bf 77}, 4872 (1996).
[arXiv:hep-ph/9606428].

\bibitem{ks97}
  A.~Kusenko and G.~Segr\`e, 
Phys.\ Lett.\ B {\bf 396}, 197 (1997).
[arXiv:hep-ph/9701311].

\bibitem{fkmp}
G.~M.~Fuller, A.~Kusenko, I.~Mocioiu, and S.~Pascoli,
Phys.\ Rev.\ D {\bf 68}, 103002 (2003). 


\bibitem{astro} See, {\it e.\,g.}, A.~G.~Lyne and D.~R.~Lorimer, Nature
  369 (1994) 127; J.~M.~Cordes and D.~F.~Chernoff,
  Astrophys. J. {\bf 505}, 315 (1998); B.~M.~S.~Hansen and E.~S.~Phinney,  
  Mon. Not. R. Astron. Soc. {\bf 291}, 569 (1997);
  C.~Fryer, A.~Burrows, and W.~Benz,  Astrophys. J. {\bf 496}, 333 (1998).

\bibitem{astro_1} 
Z.~Arzoumanian, D.~F.~Chernoff and J.~M.~Cordes,
Astrophys. J. {\bf 568}, 289 (2002).

\bibitem{explosion} I.~S.~Shklovskii, Sov. Astr., {\bf 13}, 562 (1970)
[Astr.  Zh. {\bf 46}, 715 (1970)]. 


\bibitem{fryer}
C.~L.~Fryer,
Astrophys.\ J.\  {\bf 601}, L175 (2004)
[arXiv:astro-ph/0312265].

\bibitem{sn_2D} 
A.~Burrows and J.~Hayes,
Phys. Rev. Lett. {\bf 76}, 352 (1996); 
L.~Scheck, T.~Plewa, H.~T.~Janka, K.~Kifonidis and E.~Mueller,
Phys.\ Rev.\ Lett.\  {\bf 92}, 011103 (2004).



\bibitem{drt}
O.~F.~Dorofeev, V.~N.~Rodionov and I.~M.~Ternov, Sov. Astron. Lett. {\bf
11}, 123 (1985).

\bibitem{magn} 
J.~F.~Nieves and P.~B.~Pal, Phys. Rev. {\bf D40} 1693 (1989);
J.~C.~D'Olivo, J.~F.~Nieves and P.~B.~Pal, {\it ibid.}, 3679 (1989);
S.~Esposito and G.~Capone, Z.~Phys. {\bf C70} (1996) 55;
J.~C.~D'Olivo, J.~F.~Nieves and P.~B.~Pal, Phys. Rev. Lett., {\bf 64}, 1088
(1990); 
H.~Nunokawa, V.~B.~Semikoz, A.~Yu.~Smirnov, and
J.~W.~F.~Valle, Nucl. Phys. {\bf B 501}, 17 (1997);
J.~F.~Nieves,
Phys.\ Rev.\ D {\bf 68}, 113003 (2003); 
J.~F.~Nieves,
arXiv:hep-ph/0403121.

\bibitem{phinney}
H.C.~Spruit and E.S.~Phinney, Nature, {\bf 393}, 139 (1998).

\bibitem{Fuller}
K.~Abazajian, G.~M.~Fuller and M.~Patel,
Phys.\ Rev.\ D {\bf 64}, 023501 (2001)

\bibitem{dw} S.~Dodelson and L.~M.~Widrow, Phys. Rev. Lett. {\bf 72}, 17
(1994); 
%
X.~d.~Shi and G.~M.~Fuller,
Phys.\ Rev.\ Lett.\  {\bf 82}, 2832 (1999)
A.~D.~Dolgov and S.~H.~Hansen,
Astropart.\ Phys.\  {\bf 16}, 339 (2002). 

\bibitem{eq} A.~Vilenkin, Astrophys. J. {\bf 451}, 700 (1995); 
A.~Kusenko, G.~Segr\`e, and A.~Vilenkin, Phys. Lett. B 437, 359 (1998); 
P.~Arras and D.~Lai, astro-ph/9806285.

\bibitem{voloshin} 
M.~B.~Voloshin,
Phys.\ Lett.\ B {\bf 209}, 360 (1988).
E.~Nardi and J.~I.~Zuluaga, 
Astrophys.\ J.\  {\bf 549}, 1076 (2001). 

\bibitem{farzan}
Y.~Farzan, O.~L.~G.~Peres and A.~Y.~Smirnov,
Nucl.\ Phys.\ B {\bf 612}, 59 (2001).

\bibitem{babu}
K.~S.~Babu and G.~Seidl,
arXiv:hep-ph/0312285; 
arXiv:hep-ph/0405197.

\bibitem{barkovich} 
M.~Barkovich, J.~C.~D'Olivo, R.~Montemayor and J.~F.~Zanella,
Phys.\ Rev.\ D {\bf 66}, 123005 (2002). 

\bibitem{jr}
H.~T.~Janka and G.~G.~Raffelt,
Phys.\ Rev.\ D {\bf 59}, 023005 (1999).

\bibitem{ks98} 
A.~Kusenko and G.~Segr\`e, Phys. Rev. {\bf D59}, 061302
(1999).

\bibitem{magnetic}
R.D.~Blandford, J.H.~Applegate, L.~Hernquist, 
Mon. Not. R. Astron. Soc., {\bf 204}, 1025 (1983); 

\bibitem{dt} R.~C.~Duncan and C.~Thompson, Astrophys. J. {\bf 392}, L9 
(1992); {\em ibid.}, {\bf 408}, 194.

\bibitem{magnetars} 	
C.~ Kouveliotou {\em et al.},  Astrophys. J., {\bf 510}, L115 (1999); 
R.~C.~Duncan and C.~Thompson, Bull. Amer. Astron. Soc., {\bf 29}, 140
(1997).  

\bibitem{zeldovich} Ya. B. Zeldovich, A.A. Ruzmaikin, and D.D. Sokoloff,
{\em Magnetic fields in astrophysics}, Gordon and Breach, New York, 1983.

\bibitem{aft}
K.~Abazajian, G.~M.~Fuller and W.~H.~Tucker,
Astrophys.\ J.\  {\bf 562}, 593 (2001)

\bibitem{lowT}
G.~Gelmini, S.~Palomares-Ruiz and S.~Pascoli,
arXiv:astro-ph/0403323.


\bibitem{loveridge}
L.~C.~Loveridge,
Phys.\ Rev.\ D {\bf 69}, 024008 (2004)
[arXiv:astro-ph/0309362].


\bibitem{cuesta}
H.~J.~Mosquera Cuesta,
Astrophys.\ J.\  {\bf 544}, L61 (2000); 
%
Phys.\ Rev.\ D {\bf 65}, 061503 (2002); 
H.~J.~Mosquera Cuesta and K.~Fiuza,
arXiv:astro-ph/0403529.

\bibitem{k_semi}
A.~Kusenko and D.~V.~Semikoz,
arXiv:astro-ph/0312399.


\end{thebibliography}
\end{document}